\documentclass[structabstract]{aa}  
\usepackage{graphicx}
\usepackage{txfonts}
\usepackage{color}
\usepackage{natbib}
\bibpunct{(}{)}{;}{a}{}{,} 

\begin{document}
\title{Ruling out unresolved binaries in five transitional disks}
\subtitle{VLT/NACO deep 2.12 and 1.75 $\mu$m narrow-band imaging}
\author{S\'ilvia Vicente\inst{1}, Bruno Mer\'{\i}n\inst{2}, Markus Hartung\inst{3}, Herv\'e Bouy\inst{4}, Nuria Hu\'elamo\inst{4}, \'Etienne Artigau\inst{5}, \\ Jean-Charles~Augereau\inst{6}, Ewine van Dishoeck\inst{7}, Johan Olofsson\inst{8}, Isa Oliveira\inst{7},Timo Prusti\inst{1}
\fnmsep\thanks{Based on observations collected with the NAOS-CONICA (NACO) instrument 
at the VLT in Paranal Observatory, operated by the European Southern Observatory, under programme ID 079.C-0914(A)}}

\offprints{S. Vicente, svicente@rssd.esa.int}

\institute{RSSD, European Space Agency (ESTEC), P.O. Box 299, 2200 AG Noordwijk, The Netherlands \\
\email{svicente@rssd.esa.int}
\and Herschel Science Centre, European Space Agency (ESAC), P.O. Box 78, 28691 Villanueva de la Ca\~nada (Madrid), Spain 
\and  Gemini Observatory, c/o AURA, Casilla 603, La Serena, Chile
\and CAB (INTA-CSIC), LAEFF, P.O. Box 78, 28691 Villanueva de la Ca\~nada, Madrid, Spain
\and D\'epartement de Physique and Observatoire du Mont M\'egantic, Universit\'e de Montr\'eal, C.P. 6128, Succ. Centre-Ville, Montr\'eal, QC H3C 3J7, Canada 
\and Laboratoire d'Astrophysique de Grenoble, Universit\'e Joseph Fourier, CNRS, UMR 5571, Grenoble, France 
\and Leiden Observatory, Leiden University, P.O. Box 9513, 2300 RA Leiden, The Netherlands 
\and Max Planck Institute for Astronomy, K\"onigstuhl 17, D-69117 Heidelberg, Germany 
}

\titlerunning{Ruling out unresolved binaries in five transitional disks}
\authorrunning{S\'ilvia Vicente}

\date{Received December 13, 2010; Accepted June 4, 2011}

 
  \abstract
   {
   The presence of unresolved binaries on sub-arsecond scales could explain the existence of optically thin inner holes or gaps in circumstellar disks, which are commonly referred to as ``transitional" or ``cold" disks, and it is the first scenario to check before making any other assumptions. 
      }
   {
We aim at detecting the presence of companions inside the inner hole/gap region of a sample of five well known transitional disks using spatially-resolved imaging in the near-IR with the VLT/NACO/S13 camera, which probes projected distances from the primary of typically 0.1 to 7 arcsec.
The sample includes the stars  DoAr\,21, HD\,135344B (SAO\,206462), HR\,4796A, T\,Cha, and TW\,Hya, spanning ages of less than 1 to 10 Myr, spectral types of A0 to K7, and hole/gap outer radii of 4 to 100~AU.
   }
   {
   In order to enhance the contrast and to avoid saturation at the core of the point-spread function (PSF), we use narrow-band filters at 1.75 and 2.12~$\mu$m. The ``locally optimized combination of images" (LOCI) algorithm is applied for an optimal speckle noise removal and PSF subtraction, providing an increase of~\mbox{0.5-1.5\,mag} in contrast over the classic method. 
   }
{With the proviso that we could have missed companions owing to unfavorable projections,  the VLT/NACO observations rule out the presence of unresolved companions down to an inner radius of about 0\farcs1 from the primary in all five  transitional disks and with a detection limit of  2 to 5 mag in contrast. In the disk outer regions the detection limits typically reach 8 to 9 mag in contrast and 4.7 mag for T\,Cha. Hence, the NACO images resolve part of the inner hole/gap region of all disks with the exception of TW\,Hya, for which the inner hole is only 4~AU. The 5$\sigma$ sensitivity profiles, together with a selected evolutionary model, allow to discard \emph{stellar} companions within the inner hole/gap region of T\,Cha, and down to the \emph{substellar} regime for HD\,135344B and HR\,4796A. DoAr\,21 is the only object from the sample of five disks for which the NACO images are sensitive enough for a detection of objects less massive than $\sim$ 13~M$_{\rm {Jup}}$ that is, potential giant planets or low-mass brown dwarfs at radii larger than $\sim$ 76~AU (0\farcs63).
 } 
 {These new VLT/NACO observations further constrain the origin of the inner opacity cavities to be owing to closer or lower-mass companions or other mechanisms such as giant planet formation, efficient grain growth, and photoevapo\-ration (for DoAr\,21 and HR\,4796A). 
}

 \maketitle
%

\section{Introduction}

\begin{table*}[!thb]

\caption{Observing log of the 2.12 $\mu$m VLT/NACO images of the five stars with transitional disks and corresponding PSF-calibrators}
\label{table:1}
\bigskip
\scriptsize
\centering
\begin{tabular}{lcccccccccccl}
\hline\hline
\noalign{\smallskip}
Star & Date & Grade$^{\boldmath  \mathrm{1}}$ & DIT & NDIT & Exp.time$^{\boldmath  \mathrm{2}}$ & Seeing$^{\boldmath  \mathrm{3}}$ & Airmass & $\tau_{0}^{\boldmath \mathrm{4}}$ & Strehl$^{\boldmath  \mathrm{5}}$ & FWHM$^{\boldmath  \mathrm{6}}$ &  V$^{\boldmath  \mathrm{7}}$ & Notes$^{\boldmath  \mathrm{8}}$ \\
\noalign{\smallskip}
 \& PSF &    &     & (s) &  & (min.) & (arcsec)  &  (at start) & (ms) & (\%) & (mas) &  (mag)  \\
\noalign{\smallskip}
\hline
\noalign{\smallskip}
DoAr\,21 & 2007-06-18 &  B  &10.0 & 4 &  8.0 & 0.68   & 1.530  & 2.55 & 9.3 & 129.1  & 14.01$\pm$0.03 & sat.  \\
CD-24\,13079 &  &  & 4.0 & 1, 3 & 0.73 & 0.75  &  1.547 &  2.14  &  15.2 &  85.0  & 9.86 & not sat. \\
\hline
\noalign{\smallskip}
HD\,135344B & 2007-05-09 & B & 1.0 & 12 & 2.4 & 1.75  & 1.062  & 2.73  & 13.5 & 82.5 & 8.708$\pm$0.017&    \\
HD\,136961  & & & 1.0 & 10 & 1.67 & 1.85 & 1.071 & 2.14 & 14.5 & 84.6 &   6.754 &  \\
\hline
\noalign{\smallskip}
HD\,135344B & 2007-06-20  & B & 2.0 & 6 & 2.4 & 0.77 & 1.051 & 1.07 & 25.1 & 74.4 & 8.708$\pm$0.017 & \\
HD\,144156  & & & 1.0 & 10 &  1.67 & 0.79 & 1.076 & 1.23 & 26.1 & 72.3 & 8.50 & close comp. \\
\hline
\noalign{\smallskip}
HR\,4796A & 2007-06-14 & B & 1.0 & 10 & 2.0 &  NA  & 1.069 & 1.67 & 35.9 & 69.2 & 5.78 & \\ 
HD\,109536  & & & 1.0 & 1 & 0.17 &  NA & 1.092 & 1.88 & 27.0 & 71.9 & 5.127 & \\
\hline
\noalign{\smallskip}
T\,Cha & 2007-05-23  & B & 25.0  & 2 & 10.0 & 0.85  & 1.735 & 1.37 & 4.8 & 116.2 & 11.86 & ellip. \\
CD-78\,512  &   &  & 12.0, 2.0 & 1, 6  & 1.0, 0.8  & 1.16, 1.89  & 1.739, 1.740 & 1.28,  0.76 & 7.8, 7.6  & 97.5, 97.5 &   10.14 & ellip. \\
\hline
\noalign{\smallskip}
TW\,Hya & 2007-05-12 & A & 5.0  & 5 & 5.0 & 1.46 &  1.033 & 1.27 & 19.0 & 78.4 & 11.07 & \\
HD\,94889  &  & & 1.0 & 5 & 0.83 & 1.10 & 1.053 & 0.59 & 23.9 & 75.7 & 8.80 & close comp. \\
\hline
\noalign{\smallskip}
TW\,Hya & 2007-05-23 & B & 5.0 & 5 & 5.0  & 0.98  & 1.020  & 0.94  & 18.3 & 77.9 &   11.07 & \\
HD\,101636 &  & & 0.3454 & 5 & 0.29 & 0.86 & 1.026 & 0.09 & 30.8 & 71.0 &   9.41 & \\
\hline
\end{tabular}
\begin{list} {}{}
\item[${\boldmath^{\mathrm{1}}}$] {\scriptsize the grade of each observing block (OB) is set by the on-site VLT observer based on the user-requested ambient conditions (seeing, sky transparency, airmass) vs. the actual observing conditions. A grade ``A" means that the conditions were fully within specifications, ``B" that they were mostly within specifications,  while ``C" means they were out of specifications and the OB should be repeated if possible. }
\item[${\boldmath^{\mathrm{2}}}$] {\scriptsize on source = DIT $\times$ NDIT $\times$ number of OBJECT frames}
\item[${\boldmath^{\mathrm{3}}}$] {\scriptsize average astronomical site monitor seeing at start in the visible (0.5 $\mu$m) and at the zenith (airmass=1). This value can differ significantly from the NACO image quality, which is usually better. ``NA" means that no seeing data was available for this night. } 
\item[${\boldmath^{\mathrm{4}}}$] {\scriptsize average coherence time of the atmosphere at 0.5 $\mu$m calculated by the real time computer (RTC) on real data. Short values of  $\tau_{0}$~($<$~6 ms) represent a seeing not stable in time} 
\item[${\boldmath^{\mathrm{5}}}$] {\scriptsize computed with \emph{eclipse} in the final processed image}
\item[${\boldmath^{\mathrm{6}}}$] {\scriptsize of the star in the processed image adopting a Moffat fitting to the radial profile. The {\bf full width at half-maximum (FWHM)} of the diffraction-limited PSF for the VLT at ~2.12~$\mu$m is 65~mas} 
\item[${\boldmath^{\mathrm{7}}}$] {\scriptsize optical magnitude retrieved from SIMBAD: 1) DoAr\,21 - \cite{Cieza2007}, 2) HD\,135344B - \cite{Hog2000}, 3) HR\,4796A -  \cite{Barrado2006}, 4) T\,Cha - \cite{Tanner2007}, 5) TW\,Hya - \cite{Torres2006}}
\item[${\boldmath^{\mathrm{8}}}$] {\scriptsize DoAr\,21: PSF saturated in the central pixels (4 to 7 pixel range); T\,Cha: elliptical along x-axis; PSF calibrators HD\,144156 and HD\,94889 have close companions at  d=1\farcs33, PA=20$^\circ$ and d=2\farcs2, PA=260$^\circ$,~respectively.} 
\end{list}
\end{table*}

The transition from the gas/dust-rich actively accreting optically thick disks surrounding T-Tauri and Herbig Ae/Be stars to the gas/dust-poor optically thin debris disks is currently not well understood. Of the several processes proposed for disk evolution and dissipation, the formation of planets remains the most exci\-ting \citep[e.g.,][and references therein]{Meyer2007a}. Early stu\-dies of young stars dating back to \emph{IRAS} \citep[e.g., ][]{Strom1989,Skrutskie1990} identified several sources with spectral ener\-gy distributions (SEDs) that showed a deficit of mid-infrared flux and a rise into the far-IR. These were interpreted as a sign of dust clearing, which is expected from disk evolution as the dust grows and settles or is removed with the gas by viscous accretion or other dispersal mechanisms such as photoevapo\-ration and tidal truncation because of close companions, both stellar and substellar. 
More recent surveys of young stellar clusters with {\it Spitzer} revealed a small population of  these sources in several nearby star-forming regions (d $\le$ 400 pc) based on photometry \citep{Sicilia-Aguilar2006,Muzerolle2010}, spectroscopy \citep{Brown2007,Furlan2009, Oliveira2010} or both \citep{Sicilia-Aguilar2008,Merin2010}.  They are  currently  referred to as `transitional' or `cold' disks after \cite{Brown2007},  because of their lack of emission from warm dust. The reported fraction of transitional disks can vary from a few to 50\% depending on definition and the method of classification. A consensus in nomenclature is still lacking and, as demons\-trated in \cite{Merin2010}, a significant fraction of transitional disks classified with photometry turns out to be not transitional when using spectroscopy.
The general definition of a transitional disk is a young disk  ($<$~10~Myr) with an optically thin inner region (the ``opacity hole/gap") surrounded by an optically thick outer disk (beyond the hole/gap outer radius, R$_{\mathrm {hole/gap,out}}$). 
The mid-infrared emission originates in the inner edge or  ``wall" of the truncated disk that is directly illuminated by the star \citep{Calvet2005}. Gapped disks that still retain an inner component of warm optically thick dust, typi\-cally of less than a few AU in width, are also referred to as `pre-transitional' after \cite{Espaillat2007}.  Most transitional disks are still accreting, so that gas is being transported through the inner cleared hole/gap onto the star. In some sources a small amount of micron  or sub-micron dust coexists with the gas in the hole/gap region and causes an excess over photospheric fluxes in the near-IR. 

The detection of tidal gaps in some disks has confirmed the inference of the inner holes from unresolved data. These gaps and holes have already been resolved with high-resolution contrast imaging and interferometry in the optical, infrared and at (sub)millimeter wavelengths in a handful of objects:  LkCa\,15 \citep{Thalmann2010,Pietu2006}, TW\,Hya \citep{Hughes2007}, LkH$\alpha$\,330 \citep{Brown2008,Brown2009}, GM\,Aur \citep{Dutrey2008,Hughes2009}, HR\,4796A \citep{Schneider1999,Schneider2009}, SR\,21N, WSB\,60, DoAr\,44,  HD\,135344B \citep{Andrews2009,Brown2009}, and T\,Cha \citep{Olofsson2011}. The direct images and SEDs modeling of transitional disks suggests inner cavities in the dust distribution of  3 to 100 AU in extension \cite[e.g.,][]{Dalessio2005b,Najita2007,Brown2007,Cieza2010,Olofsson2011}. 
 Although se\-ve\-ral mechanisms have been proposed to explain the origin of inner disk holes or cleared gaps,  the exis\-tence of previous\-ly unresolved close binaries has been confirmed in se\-ve\-ral transitional disks  \citep{Ireland2008,Pott2010} and is the first scenario to check before making any other assumptions.

This paper presents near-infrared high-spatial resolution narrow-band imaging of five well known transitional disks  --  DoAr\,21, HD\,135344B (SAO\,206462), HR\,4796A, T\,Cha, and TW\,Hya --  and is aimed at detecting close companions within their inner hole/gap region.
The VLT/NACO observations and data reduction are presented in \S~\ref{observations}. The results, including the PSF-subtracted images and the sensitivity curves are shown in \S~\ref{results}, where we also refer to the mass limits and orbit cons\-traints of possible companions. In \S~\ref{discussion} we tentatively classify the five disks on basis of the mechanisms that potentially create their inner opacity hole/gap.  We also ana\-ly\-ze the results for each target independently and discuss how the constraints derived from the NACO images complement or support previous studies of companions that made use of other techniques. 
The conclusions of this study are given in \S~\ref{conclusions}.


\section{Observations and data reduction}
\label{observations}

\subsection{Observations}
The target stars were observed with NACO, an adaptive optics supported
high-resolution imager and spectrograph installed on ESO's Very Large
Telescope in Chile \citep{Rousset2003,Lenzen2003}.  The observations
were performed in service mode in several runs between May and June
2007 with two narrow-band filters, NB\_2.12 (2.122 $\pm$ 0.011
$\mu$m) and NB\_1.75 (1.748 $\pm$ 0.013 $\mu$m), to avoid saturation
in the core of the PSF.  The images were collected with the highest
spatial re\-so\-lution camera S13 (13.27 mas/pixel,
14$\arcsec$\,$\times$\,14$\arcsec$ FoV) and with the NAOS visible
wavefront-sensor and the vi\-si\-ble dichroic. The camera `Double
RdRsRd' readout mode and the `HighDynamic' mode (15\,000 ADU Full Well
Depth) were selected to observe these bright stars.  The PSF standard
stars were chosen to be close in spectral type and parallactic angle
to the target stars. They were observed immediately after the science
targets to match sky conditions as well as possible.  More details on
the observations can be found in Table~\ref{table:1}.  The observing
conditions of the different runs were mostly poor (bad and highly
variable seeing, clouds, winds), resulting in a low adaptive optics
(AO) performance.  The analyzed datasets barely met the required
specifications; part of the data had to be discarded. Only one dataset
was fully within specifications. Hence, future datasets obtained under
more favorable atmospheric conditions will certainly further improve
the results.

\begin{figure}[!t]
\centering
\includegraphics[width=8cm]{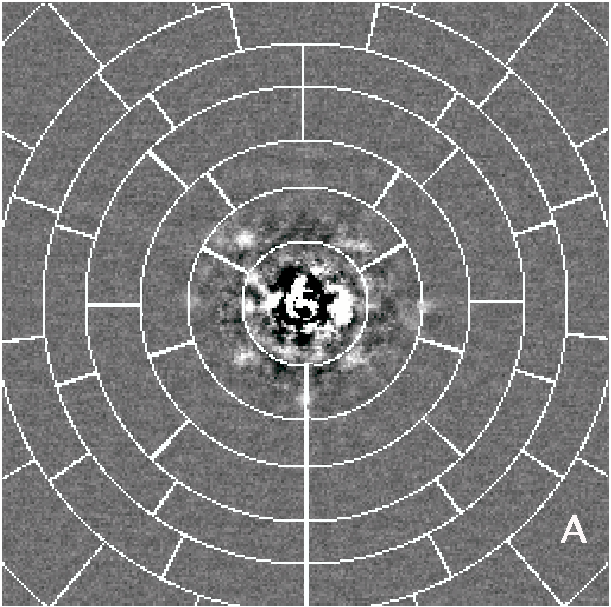}
\caption{LOCI optimization grid for a full width at half-maximum (FWHM) of 6~pixels overlaid on the 1.75 $\mu$m PSF-subtracted image of HD\,135344B (09-05-2007).  The frame is 512 $\times$ 512 data pixels (13.27~mas/pixel) and the diameter of the inner circle is $\sim$~52~pi\-xels. 
The FWHM input used in the LOCI pipeline was adjusted to the average FWHM of the data for each target (for most of the data FWHM=6~pixels and for some FWHM=9 and 10~pixels). The rele\-vant input  parameters for LOCI are not the exact annulus sizes but the surface area of the ring subsections $A$ (see explanation in section~\ref{LOCI}), which in our case corresponds to 70 times the PSF area. The algorithm adjusts the ring sizes to match this area condition (70 PFS cores/segment) and consequently the ring sizes change slightly for different FWHMs and for different radial positions (because the farther out, the more segments). }
\label{fig:LOCI_grid}
\end{figure}

\subsection{Data reduction and the \emph{LOCI} algorithm}
\label{LOCI}
We adopt a strategy different from the ``classic'' PSF subtraction
approach that takes advantage of the stability of the quasi-static
speckle pattern of the AO-corrected PSF.  Each indivi\-dual image
(DIT) of the target is dark-subtracted and flat-fielded \mbox{using}
standard reduction procedures with IRAF and the NACO pipeline, which is 
based on the \emph{eclipse} library \citep{Devillard1997}.  
For each of them, the stellar PSF speckles are attenuated by
subtracting an optimized PSF obtained using the ``locally optimized
combination of images" (LOCI) algorithm detailed in \cite{Lafreniere2007}.
 Conceptually, this data processing technique uses the fact
that high-contrast imaging PSFs from a given instrument - even when
taken with different observational setups - share common speckle
structures. If an observing strategy is arranged in a way that a set of
images with similar speckle patterns can be taken while a faint
companion is present in one image and absent in others, one can build
a `best estimate' of the PSF for a given image from the set of
reference images (PSF images). This `best estimate' is subtracted from
the target frame and a significant gain in sensitivity can be achieved. 

This approach was initially conceived in the context of angular
differential imaging or ADI (see demonstration in Figure~8 of \citealt{Lafreniere2007}
 and its comparison with the \citealt{Marois2006} ADI results),
where a long sequence of observations is taken with the Cassegrain
rotator turned `off'. Through the sequence, a companion at a given
field position will rotate through the field as the parallactic angle
changes but the average speckle pattern caused by telescope and
ins\-trument aberrations typically remains stable.  With this sequence,
a high-precision PSF frame can be built for {\it each} image frame
from all other frames taken at sufficiently different angular
positions (to avoid self-subtraction of a companion).
This technique can be extended to any set of images where a putative
companion is absent or significantly suppressed\footnote{\scriptsize E.g. by
spectral differential imaging (ADI). See in \cite{Artigau2008} the application of LOCI to
the Gemini Near-Infrared Coronographic Imager (NICI) data, a dual-channel ima\-ger exploiting the methane absorption feature for planet search.}. That means it can
also be applied to a conventional data set (field rotation compensated for by the Cassegrain rotator following) using PSF stars that are separately recorded as we do in this paper.

 \begin{figure*}[!th]
\centering 
\includegraphics[width=17cm, keepaspectratio]{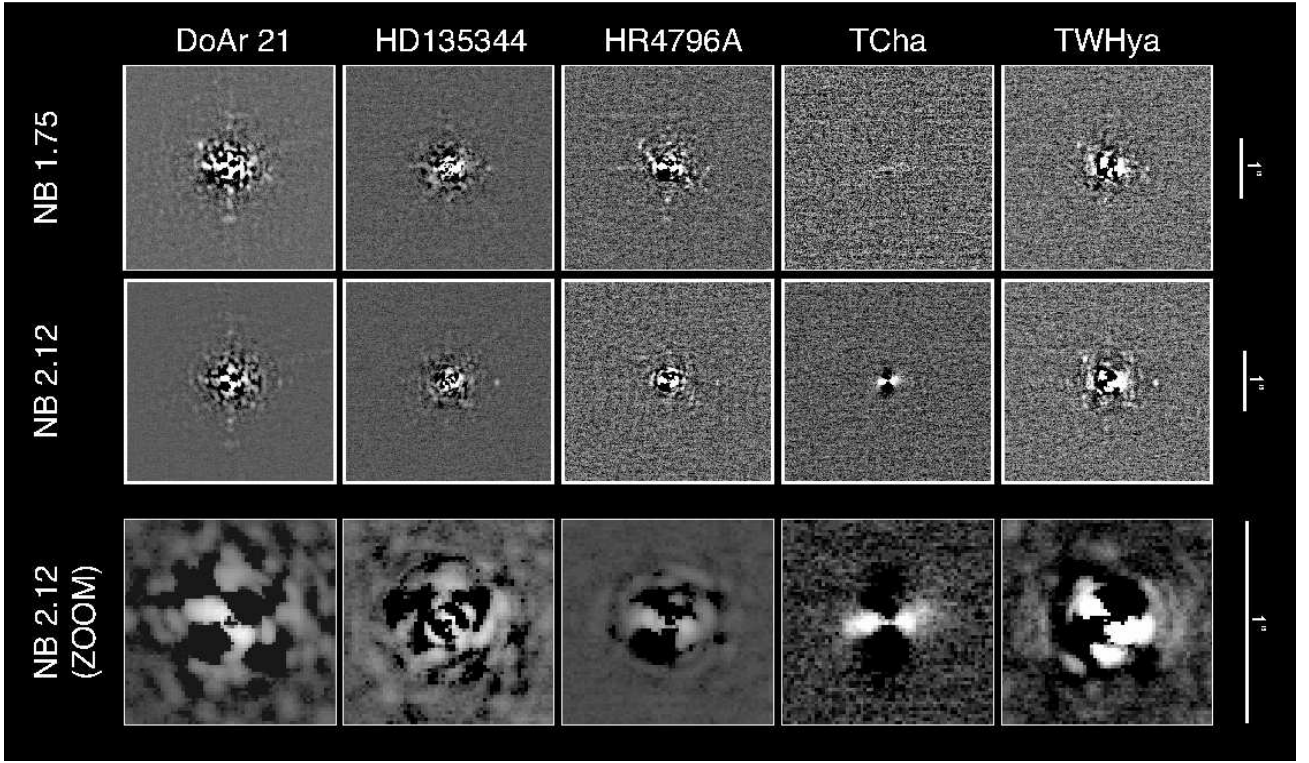}
\caption{High-contrast residual images after PSF removal with the LOCI algorithm for both
  filters NB\_1.75 and NB\_2.12. Each panel is a
  $3\farcs3\times3\farcs3$ sub-image centered on the target star.  The bottom set of panels is a zoom-in of the inner $1''\times1''$ region for the NB\_2.12 filter showing the PSF residuals at small distances from the primary. 
 No speckles were found to be achromatic (i.e., no evidence for a real object)  and present in the two epochs of HD\,135344B ($\sim$1.5 month apart) and TW\,Hya (just a few days apart) for which only the best dataset is displayed. 
  North is up and east to the left.}
\label{fig:LOCI_results}
\end{figure*}

The heart of the LOCI algorithm can be described quite easi\-ly: a specific target frame is divided into ``optimization
subsections" and the best linear combination of reference images is found for every optimization subsection to represent the corres\-ponding target frame section. The idea to split the target ima\-ge
into subsections takes into account that the correlation between
target and reference (PSF) ima\-ges typically varies with its field position.
A crucial step in the LOCI algorithm is the definition of a proper set
of reference ima\-ges.  In angular differential ima\-ging data for e.g., the displacement caused by field rotation has to be large enough to avoid self-cancellation. This
complication does not exist in our case.  We simply used ano\-ther set of
PSF star observations with an identical instrumental setup.  But we exploit
the fact that NACO PSFs show a strong degree of mirror symmetry and 
also add the mirror-flipped versions of each PSF frame to the set of
refe\-ren\-ce frames\footnote{\scriptsize Note that obviously LOCI provides a higher
gain in sensitivity for ADI data because there is no additional slew to
acquire the PSF star and furthermore the pupil does not rotate. Therefore, there is a 
better match of the quasi-static speckle pattern for the reference and target frames.}.  
For the NACO data we adjusted the surface area of the optimization
subsections $A$ (as seen in the grid in Figure~\ref{fig:LOCI_grid}) to contain approximately 70 ``PSF cores", i.e., we set Lafreni\`ere's parameter $N_A$ to 70 in Equation~1 of \cite{Lafreniere2007}:

 \begin{equation} \label{eq:1} \\
  A = N_A \pi \left( \frac{FWHM}{2} \right)^2, 
 \end{equation} 

\noindent where the FWHM is the full width at half-maximum of the PSF and $N_A$ the number of 
``PSF cores" that fit in the optimization subsection.
We did not pursue an extensive optimization for this parameter but kept it large enough for convenient
calculation times and to be on the safe side concerning self-cancelation.  Figure~\ref{fig:LOCI_grid} shows the annulus sizes and geometry of the optimization subsections $A$ for a FWHM of 6 pixels.

After subtracting the reconstructed PSF from every
frame, the images were registered and median-combined. The remai\-ning
speckles add up incoherently in the final ima\-ge, while any companion
would add up coherently. The final images or PSF-subtraction residuals are shown in Figure~\ref{fig:LOCI_results}. 


\section{Results}
\label{results}

\begin{figure*}[th]
\centering
    \includegraphics[width=17cm]{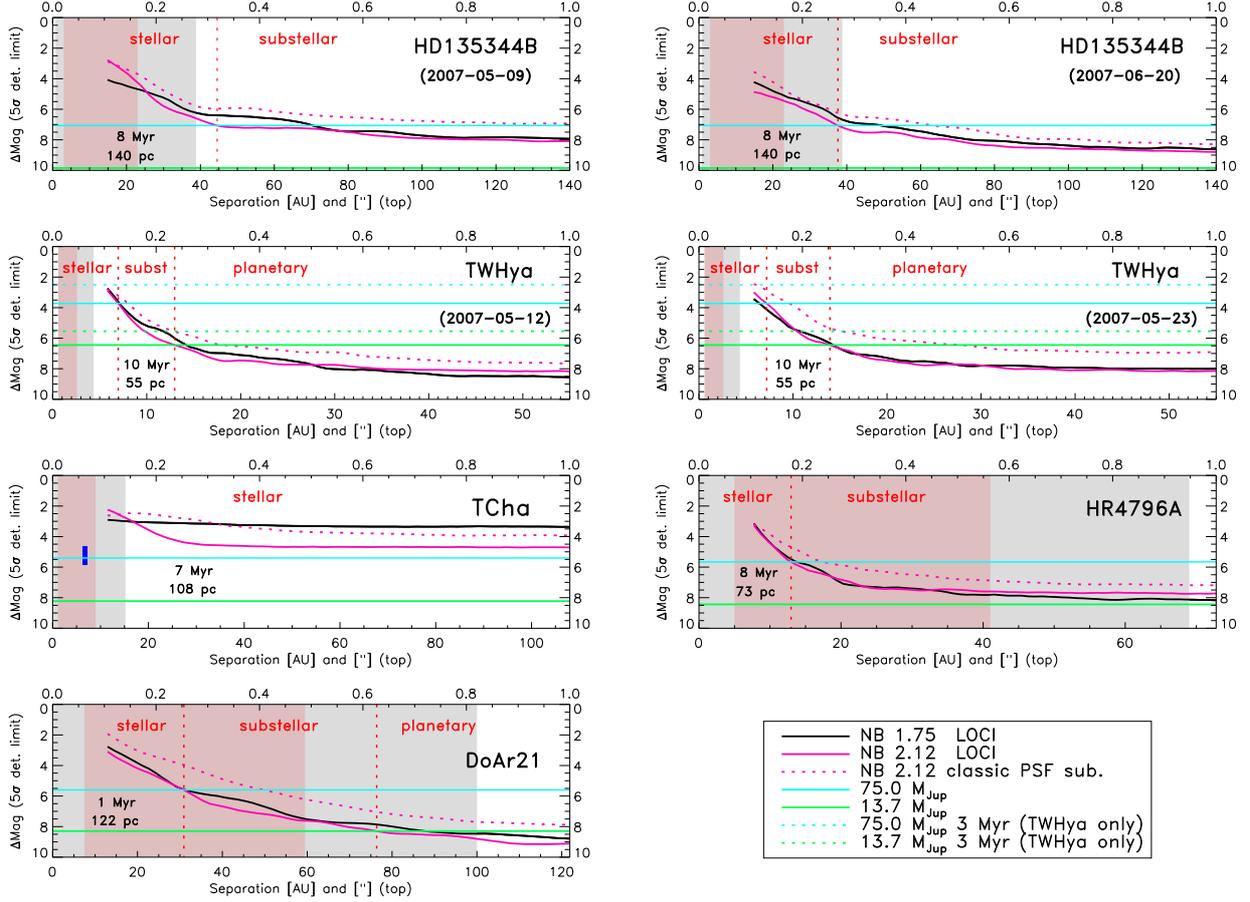}
\caption{$5\sigma$-sensitivity profiles as a function of the distance from
  the star. The NB\_1.75 and NB\_2.12 filters are shown in black and
  pink lines, respectively. For comparison, the dashed curve shows the
  $5\sigma$-sensitivity for standard PSF-subtracted  images at 2.12 $\mu$m, not using
  LOCI. The cyan and green lines mark the approximate positions of 75
  and 13.7 M$_{\rm Jup}$, the canonical values that define the
  substellar and planetary regime boundaries according to the
  BT-DUSTY models from \cite{Allard2010}. The dashed lines represent the same boundaries but for an age of  3 Myr for TW\,Hya recently reported by \cite{Vacca2011}. 
  The gray areas show the disk inner hole/gap sizes reported in the literature and the red dashed vertical lines mark the 
   intersections of the LOCI contrast curve at 2.12 $\mu$m with the cyan and green lines.
   They indicate the radial distance from the primary from where the NACO images are sensitive 
   to substellar and planetary mass objects according to the adopted BT-DUSTY model.  The pink areas represent the hole/gap distance range where stellar companions could exist according to the models of \cite{Artymowicz1994} on basis of the hole/gap size.  For typical gaseous viscous disk parameters, and $a$ denoting the binary \mbox{semi-major axis}, the inner edge location of the circumbinary disk varies from $1.7a$ to $3.3a$ for binary eccentricity increasing from 0 (circular orbit) to 0.75, and assuming a face-on geometry ($i$= 0$^{\circ}$, $PA$= 0$^{\circ}$). The position of the companion varies from $a(1-e)$ and $a(1+e)$, the periapsis and apoapsis of the secondary relative to the primary.  For T\,Cha, the NACO images are only sensitive to stellar mass objects brighter than 4.7 mag in contrast or 0.12~M$_\odot$. The blue vertical bar in the T\,Cha plot indicate the approximate 5-$\sigma$  detection limit ($\Delta K_s>4.65$~mag or mass $<$~0.12~M$_\odot$) at the separation ($6.7\pm1$~AU) and position angle ($\sim 78 \pm 1$ deg) of the companion candidate detected in the $L'$-band with NACO by \cite{Huelamo2011}. A lower limit of 50~M$_{\rm {Jup}}$ was chosen following Fig.~4b in their paper. This detection is  well within the disk hole/gap of T\,Cha and inside the binary range predicted by \cite{Artymowicz1994}. 
 The profiles are displayed between  0.1 and 1 arcsec for all stars.
  Simulations with a fake target of 4 mag in contrast show that the sensitivity curves are reliable as close as 100-120 mas from  the primary (between the first and second Airy ring at 2.12 $\mu$m). Beyond 1 arcsec and up to 3.3 arcsec the contrast curves remain flat.
 \label{fig:Sensitivity}}
\end{figure*}

 Figure~\ref{fig:LOCI_results} shows the final images after the data reduction and 
analysis with the LOCI high-contrast technique described above. Each tile ($3\farcs3\times3\farcs3$)
is centered on the target star and each row displays the residual images in one of the two narrow-band filters.
The bottom set of panels is a zoom-in of the inner $1''\times1''$ region for the NB\_2.12 filter.
The optimal suppression of speckle noise and the improvement in the sensitivity limit allows us 
to search for eventual compa\-ni\-ons.  
Some speckles remain in the images after LOCI and are caused by the non-optimal observational strategy.
Electronic ghosts aligned with the X and Y axis and row saturation effects typical of NACO were present in many of the processed frames of the target stars and PSF-calibrators.
These residuals can be discarded by blinking between the 1.75 and 2.12 $\mu$m images, or between different datasets if they were acquired close in time and hence are not affected by the orbital motion of a possible companion. 
A real compa\-nion looks like a speckle that does not move between the different filters (achromatic) or different datasets when available, as in the case of TW\,Hya (just a few days difference) and HD\,135344B ($\sim$1.5 month).
A careful visual analysis of the images discarded any achromatic speckles  
that could point to a tentative detection of nearby sources. 
The best images of HD\,135344B (20~June~2007) and TW\,Hya  (12~May~2007) are shown.  
 We note that even though our data are not ideal for LOCI application (poor AO performance, low contrast,  and speckles are less dominant), the results are still significantly better than those obtained with ``classic" PSF subtraction (see Figure~\ref{fig:Sensitivity})

\begin{table*}[!thb]

\caption{Physical properties of the five stars and their disks found in the literature}
\label{table:2}
\bigskip
\scriptsize
\centering
\begin{tabular}{lcccccccc}
\hline\hline
\noalign{\smallskip}
Star & SpT & $K_{\rm s}$${\boldmath^{ \mathrm{1}}}$ & $\Delta K_s$${\boldmath^{ \mathrm{1}}}$ & Dist. & $A_V$ & M$_{\rm star}$ & Age & $\dot M_{\rm acc}$${\boldmath^{ \mathrm{2}}}$  \\
\noalign{\smallskip}
  &  &  & & (pc) &  & (M$_\odot$) & (Myr) & (M$_\odot \mathrm {yr}^{-1}$)\\
\noalign{\smallskip}
\hline
\noalign{\smallskip}
DoAr\,21 & K1${\boldmath^{ \mathrm{a}}}$ & 6.227  & 0.018 & 121.9$\pm$5.8${\boldmath^{ \mathrm{b}}}$  & 6.2${\boldmath^{ \mathrm{a}}}$ & 1.8${\boldmath^{ \mathrm{a}}}$ & $<$1${\boldmath^{ \mathrm{a}}}$ & $<$ 10$^{\mathrm {-11, c}}$ \\
HD\,135344B & F4${\boldmath^{ \mathrm{e}}}$ & 5.843  & 0.020 &140$\pm$42${\boldmath^{ \mathrm{g}}}$ & 0.5${\boldmath^{ \mathrm{e}}}$ & 1.7$\pm$0.2${\boldmath^{ \mathrm{g}}}$ & 8$\pm$4${\boldmath^{ \mathrm{g}}}$ & 5$\times10^{\mathrm {-9,g}}$  \\
HR\,4796A &  A0${\boldmath^{ \mathrm{i}}}$ & 5.769 & 0.023 & 72.8$\pm$1.7${\boldmath^{ \mathrm{h}}}$ & -- & 2.5${\boldmath^{ \mathrm{i}}}$ & 8$\pm$2${\boldmath^{ \mathrm{i}}}$ & --    \\ 
T\,Cha${\boldmath^{ \mathrm{6}}}$ & G8${\boldmath^{ \mathrm{j,k}}}$ & 6.954 & 0.018 & 108$\pm$9${\boldmath^{ \mathrm{l}}}$ & 1.7${\boldmath^{ \mathrm{k,5}}}$ & 1.5${\boldmath^{ \mathrm{j}}}$ & 7${\boldmath^{ \mathrm{k,5}}}$ &  4$\times10^{\mathrm {-9,k}}$        \\
TW\,Hya${\boldmath^{ \mathrm{7}}}$ & K7${\boldmath^{ \mathrm{o}}}$ & 7.297  & 0.024 & 55$\pm$9${\boldmath^{ \mathrm{h}}}$ & 0.0${\boldmath^{ \mathrm{q}}}$ & 0.6${\boldmath^{ \mathrm{r}}}$ & 10${\boldmath^{ \mathrm{p}}}$, 3${\boldmath^{ \mathrm{*}}}$ & 8$\times$10$^{\boldmath^{ \mathrm{-10,q}}}$-- 10$^{\boldmath^{ \mathrm{-9,y}}}$  \\
\noalign{\smallskip}
\hline\hline
\noalign{\smallskip}
Disk &  R$_{\rm disk,in}$${\boldmath^{\mathrm{3}}}$  & R$_{\rm hole/gap,in}$${\boldmath^{\mathrm{3}}}$ & R$_{\rm hole/gap,out}$${\boldmath^{\mathrm{3}}}$ &  R$_{\rm disk,out}$${\boldmath^{\mathrm{3}}}$  & $i$ & PA & M$_{\rm disk}$${\boldmath^{\mathrm{4}}}$  & Class \& disk type${\boldmath^{ \mathrm{5}}}$  \\
\noalign{\smallskip}
  & SED, obs. & SED, obs. & SED, obs. & SED, obs. & \\
\noalign{\smallskip}
 & (AU) & (AU) & (AU) & (AU) & (deg) & (deg) & (M$_\odot$) \\
 \noalign{\smallskip}
\hline
\noalign{\smallskip}
DoAr\,21 &  -- & -- &  --, 100${\boldmath^{ \mathrm{a}}}$ & -- &  $<$ 45${\boldmath^{ \mathrm{d}}}$ & -- &  $<$ 0.001${\boldmath^{ \mathrm{c}}}$ &PMS~~circumbinary/debris${\boldmath^{ \mathrm{c}}}$  \\
HD\,135344B & 0.18${\boldmath^{ \mathrm{e}}}$, 0.05$\pm$0.25${\boldmath^{ \mathrm{w}}}$ & 0.45${\boldmath^{ \mathrm{e}}}$, 1.8$\pm$0.2${\boldmath^{ \mathrm{w}}}$  & 45${\boldmath^{ \mathrm{e}}}$, 39$\pm$4${\boldmath^{ \mathrm{f}}}$  & 300${\boldmath^{ \mathrm{e}}}$, 160${\boldmath^{ \mathrm{g}}}$  & 21${\boldmath^{ \mathrm{f}}}$ & 55${\boldmath^{ \mathrm{f}}}$ & (2.8$\pm$1.3)$\times$10$^{\mathrm {-3,g}}$ & PMS~Herbig${\boldmath^{ \mathrm{~g}}}$\\
HR\,4796A & -- & -- & 40-200${\boldmath^{ \mathrm{i}}}$, 69${\boldmath^{ \mathrm{i}}}$  & --, 87${\boldmath^{ \mathrm{i}}}$  & 75.88$\pm$0.16${\boldmath^{ \mathrm{i}}}$ & 27.01$\pm$0.16${\boldmath^{ \mathrm{i}}}$ & $\ge$ 7.4$\times$10$^{-5,}$${\boldmath^{ \mathrm{i}}}$ & Herbig~~debris disk${\boldmath^{ \mathrm{i}}}$ \\ 
T\,Cha${\boldmath^{ \mathrm{6}}}$ & 0.08${\boldmath^{ \mathrm{e}}}$ 0.13${\boldmath^{ \mathrm{n}}}$, --  &   0.2${\boldmath^{ \mathrm{e}}}$ 0.17${\boldmath^{ \mathrm{n}}}$, -- & 15${\boldmath^{ \mathrm{e}}}$ 7.5${\boldmath^{ \mathrm{n}}}$, --  & 300${\boldmath^{ \mathrm{e,n}}}$, -- & 75${\boldmath^{ \mathrm{j}}}$, 60${\boldmath^{ \mathrm{n}}}$ & 78$_{-50}^{ \mathrm{+87,n}}$ & 0.003${\boldmath^{ \mathrm{m}}}$, (1.76$\pm$0.25)$\times$10$^{\mathrm{-2,n}}$ & WTT${\boldmath^{ \mathrm{z}}}$\\
TW\,Hya${\boldmath^{ \mathrm{7}}}$ & $\sim$0.02${\boldmath^{ \mathrm{r}}}$, -- & 0.25${\boldmath^{ \mathrm{t}}}$, 0.06${\boldmath^{ \mathrm{u}}}$ & 4${\boldmath^{ \mathrm{r}}}$, 4.3$\pm$0.3${\boldmath^{ \mathrm{s}}}$  & 230${\boldmath^{ \mathrm{x}}}$ &  4.3$\pm$1.0${\boldmath^{ \mathrm{v,1}}}$, 7$\pm$1${\boldmath^{ \mathrm{v,2}}}$  & 332$\pm$10${\boldmath^{ \mathrm{v}}}$ &  $\sim$0.06${\boldmath^{ \mathrm{r}}}$ & CTT${\boldmath^{ \mathrm{*}}}$ \\
\noalign{\smallskip}
\hline
\end{tabular}
\begin{list} {}{}
\item[${\boldmath^{\mathrm{1}}}$] {\scriptsize from 2MASS}
\item[${\boldmath^{\mathrm{2}}}$] {\scriptsize average accretion rate derived from H$\alpha$ line emission for DoAr\,21 and T\,Cha, Br$\gamma$ line emission for TW\,Hya~(y) and HD\,135344B~(g), and excess Balmer continuum for TW\,Hya~(q)}
\item[${\boldmath^{\mathrm{3}}}$] {\scriptsize radius from the primary derived from both SED fitting models and spatially resolved observations in the optical to the sub-mm/mm. In this work we adopt the latter values when available}
\item[${\boldmath^{\mathrm{4}}}$] {\scriptsize total disk masses estimated from mm fluxes assuming a gas-to-dust ratio of 100 or 77  for TW\,Hya. T\,Cha: the first value is that of the 3 mm emission flux for a distance of 66~pc. Because M$_{\rm disk}\propto$~d$^2$, a value of 0.008 M$_\odot$ is expected for the assumed distance d=108 pc. The second value was derived from model fitting to the observed SED and $H$ and $K$-band interferometric observations with the VLTI/AMBER instrument} 
\item[${\boldmath^{\mathrm{5}}}$] {\scriptsize young stellar object (YSO) evolutionary class and disk type found in the literature: pre-main sequence (PMS), weak lined T-Tauri (WTT) and classical T-Tauri (CTT) star}
\item[${\boldmath^{\mathrm{6}}}$] {\scriptsize T\,Cha shows variable circumstellar extinction Av=1.2 -- 4.6 and an age of 4.1-10 Myr \citep[and references therein]{Schisano2009}. Here we adopt the most frequent value Av= 1.7 and age of 7 Myr.  The disk inner hole size adopted is from \cite{Brown2007}  }
\item[${\boldmath^{\mathrm{7}}}$] {\scriptsize a recent work by *\cite{Vacca2011} based on near-infrared spectroscopy revises the spectral type and age of TW\,Hya to M2.5 (instead of K7) and $\sim$ 3 Myr (instead of 8--10~Myr). \cite{Akeson2011} combines new near-IR interferometric data with previous spatially resolved observations at 10 $\mu$m and 7 mm to constrain disk models based on a flared disk structure. They find TW\,Hya fits a three-component model composed by an optically thin emission from $\sim$ 0.02 AU to $\sim$ 0.5 AU,  an  optically thick $\sim$ 0.1 AU wide ring that is followed by an opacity gap, and an outer optically thick disk starting at 3.8 -- 4.5 AU. Here we adopt the values of 0.5~AU and 4.3~AU for the inner and outer hole/gap radius, respectively}
\item[]{\scriptsize} 
\item[]{{\bf References.} a)~\cite{Jensen2009}, b)~\cite{Loinard2008}, c)~\cite{Cieza2010,Cieza2008},  d)~\cite{Bary2002}, e)~\cite{Brown2007}, f)~\cite{Brown2009}, g)~\cite{Grady2009} and references therein, h)~\cite{vanLeeuwen2007}, i)~\cite{Schneider1999,Schneider2009}, \cite{Jura1995}, \cite{Besla2007} and references therein,  j)~\cite{Alcala1993}, 11k)~\cite{Schisano2009}, l)~\cite{Torres2008}, m)~\cite{Lommen2007}, n)~\cite{Olofsson2011}, o)~\cite{Herbig1978}, p)~\cite{Webb1999}, q)~\cite{Herczeg2009}, r)~\cite{Calvet2002}, s)~\cite{Hughes2007}, t)~\cite{Thi2010}, u)~\cite{Eisner2006},  v,1)~\cite{Pontoppidan2008} for the inner disk, v,2)~\cite{Qi2004} for the outer disk, w)~\cite{Fedele2008}}, x)~\cite{Weinberger2002}, y)~\cite{Eisner2010}, z)~\cite{Alcala1993}
\end{list}
\end{table*}

\subsection{Detection and mass limits for putative stellar/substellar companions} 

In order to quantify the upper limits on mass of potential
companions, we compute the differential magnitude profile as a
function of the separation from the primary star. They are obtained by
calculating the $5\sigma$ RMS over concentric annuli (1~FWHM width) of the intensity
on the residual maps (PSF-subtracted) displayed in Figure~\ref{fig:LOCI_results}.
These resulting sensitivity curves are shown for each filter and image
in Figure~\ref{fig:Sensitivity}.  The black and pink lines represent
the differential magnitude curves for the \mbox{NB\_1.75} and \mbox{NB\_2.12}
filters, respectively. For comparison, the dashed curve shows the
$5\sigma$-sensitivity not applying LOCI but standard PSF
subtraction to the 2.12~$\mu$m i\-ma\-ges. 
The LOCI algorithm improves the detection sensitivity by  0.5 up to 1.5 mag which corresponds to
an improvement of up to factor of 4 in contrast ratio.
The projected se\-pa\-ration in AU was computed using
the distance to the stars from Table~\ref{table:2} and is displayed at
the bottom x-axis. The separation in arcsec is shown at the top
x-axis.
We typically reach a contrast of 2 to 5 mag at  0\farcs1 and 4 to 6 mag at 0\farcs2.  The
mass limits can be estimated with a selected
evolutio\-na\-ry model of a substellar object.  The cyan and green
lines mark the approximate positions of the ca\-no\-ni\-cal~va\-lues
of 75 and 13.7~M$_{\rm Jup}$ \citep{Burrows1997,Burrows2001} that
define the substellar and plane\-tary regime boundaries\footnote{
  \scriptsize{The mass boundaries between stars, brown dwarfs, and
    planets are model-dependent. The former is defined as the minimum
    mass required for burning hydrogen for a solar metallicity object
    in the main-sequence. This value is a function of the helium
    fraction, metallicity and opacity of grains and, depending on the
    assumed parameters, it can range from 0.07 to 0.092 M$_{\odot}$ or
    73 to 96 M$_{\rm Jup}$ \citep[and references
      therein]{Burrows1997,Burrows2001}.  The latter is defined as the
    mass-limit for deuterium burning and varies with the helium
    abundance, the initial deuterium abundance, the metallicity and
    the fraction of the initial deuterium abundance that must combust for the substellar object to qualify as having burned
    deuterium \citep{Spiegel2010}. The deuterium mass limit can range
    from $\sim$ 11.0 to $\sim$ 16.3~M$_{\rm Jup}$ depending on the model
    assumptions.  A final remark is that other criteria, such as the
    formation process, should be considered when classifying a
    substellar object.}} for comparison with our data.  They were
computed by interpolating the mass vs. \mbox{NB\_ 2.12} relation of the
BT-DUSTY models from \mbox{\cite{Allard2010}}\footnote{\scriptsize {These are the models of brown dwarfs and very
    low-mass stars with dusty atmospheres from \cite{Allard2001} and
    \cite{Chabrier2000} but with updated opacities. One should be cautious when applying evolutionary models developed for stars and substellar objects to derive the mass of planets. The mass of a freshly formed planet only depends on the accretion history and accretion rate and a planetary model should be considered instead.}}
convolved with the NACO filter at the distances and ages listed in  Table~\ref{table:2}. We then subtracted the 2MASS $K_{\rm s}$
absolute magnitudes of the stars (at the assumed distance and corrected for extinction  using the interstellar extinction law of \citealt{Rieke1985} with $R_V$= 3.1) to obtain the difference in magnitude expected
for a companion of 13.7 and 75~M$_{\rm Jup}$. The NB\_ 2.12 filter
was selected because a better contrast is expected at longer
wavelengths. The approximation of \mbox{NB\_2.12}~mag $\sim$ $K_{\rm s}$ mag
is a good first estimate (within $\le$~0.1~mag) that does not
compromise the results if we account for all the errors involved.
A younger age for TW\,Hya of 3 Myr (instead of 8-10 Myr) has been reported in \cite{Vacca2011}. We computed the substellar and planetary mass boundaries of TW\,Hya for an age of 3 Myr and plotted it as cyan and green dashed lines for comparison. For this case, the NACO images are sensitive to substellar mass objects within the whole distance range probed (0\farcs1 to 7\farcs0 from the star). 
The gray area in the plots represents the size of the inner hole/gap
determined from previous stu\-dies (Table~\ref{table:2}). Three of the five disks in the sample have an optically thick inner ring of sub- to nearly 2~AU in radii reported in the literature. Here we adopt the inner hole/gap radii of 1.8 AU for  HD\,135344B \citep{Fedele2008}, 0.2 AU for T\,Cha \citep{Brown2007} and 0.6 AU for TW\,Hya from the recent results of \cite{Akeson2011}. Nevertheless, these values are not relevant for the analysis in this paper because the NACO observations are not sensitive to these inner radii.
The vertical  red dashed lines give an indication of the disk distance range where the images are sensitive 
to the substellar and planetary regimes, according to the BT-DUSTY model adopted here.
The inner 100~mas of the profiles is omitted. The most inner central  region of the residual maps is not dominated anymore by speckle or other noise but by systematic errors in removing the PSF core. In this region the ``standard" contrast curve calculation does not apply anymore. Simulations with a 4  mag contrast fake target show that the contrast curves are reliable to about the inner 100-120 mas or between the first and second Airy ring at~2.12~$\mu$m.
Table~\ref{table:2} lists the physical parameters of the five stars and their disks found in the literature.

\subsection{Orbit constraints on putative stellar/planetary companions} \label{sec:orbit}

This paper addresses  one possible origin for inner holes or cleared gaps in protoplanetary disks, which is 
the dynamical clearing by unresolved companions.  In this section, we discuss the expected orbits of potential companions that account for the observed hole/gap size on basis of existing models or numeri\-cal simulations for both planets  (\citealt{Quillen2004,Varniere2006b}) and stars \citep{Artymowicz1994}. 

We caution that these quantitative estimates are very uncertain because we do not know the orbital parame\-ters and only poorly constrain the disk inclinations and other proper\-ties. The important result in this paper is qualitative and refers to the pre\-sence or absence of stellar/substellar compa\-nions beyond the 0\farcs1 radius  probed with the NACO images. We cannot completely rule out  companions as long as there is a small chance of alignment during the observations that could have occulted the companion. But the likelihood of that happening is small.
Table~3 shows the results for the distance range around
each of the stars where we can effectively rule out the presen\-ce
of stellar and substellar companions on the basis of our NACO
observations and with the limitation that we could have missed companions because of unfavorable projections.

\subsection*{\bf Planetary companions}

A single planet of a few M$_\mathrm {Jup}$ carves only a narrow gap (typi\-cally $\Delta r_\mathrm {gap}/r_\mathrm {gap} \sim$ 0.1) about its orbit in a viscous disk that is a few times more massive than the planet and as soon as the Hill's  radius\footnote { \scriptsize {The Hill radius of a planet is  given by $R_H = a_p (1-e) (\frac{m_p}{3M_{\star}})^{1/3}$, where $a_p$ is the semi-major axis of the orbit of the planet, $e$ the eccentricity, $m_p$ the mass of the planet and $M_\star$ the mass of the star.}} is larger than the disk height  \cite[e.g.,][]{Lubow2006}. 
This is not compatible with the large inner hole/gap observed in most transitional disks. Nevertheless, a system of multiple plane\-ts could open large holes and maintain sharp inner disk edges \cite[and refe\-rences therein]{Perez-Becker2011}. Another possibility is to consider that the mass of the single planet is large compared to the disk mass with which it interacts so that the inertia of the planet slows the migration and the inner disk within the planet's semi-major axis has time to viscously accrete onto the star, creating a large inner hole \citep{Quillen2004}. The material beyond the planet's orbital radius is prevented from accreting by the transfer of orbital angular momentum from the planet to the disk and piles up, forming the inner rim of the outer disk. 
Additionally, if the massive planet  forms faster than the migration timescale and slower than the hole/gap at its formation radii, numerical si\-mulations by  \cite{Varniere2006b} show that the ``spectral hole" always accompanies the planet and extends from the planet's orbital radius to the stellar surface.  Here ``spectral hole" subtends a region with reduced density, which is not necessarily fully depleted of material, allowing for accretion to hold even after the hole has formed. 
If the planet forms faster than the hole formation timescale, a gap can open with an inner radius larger than the star radius and an outer edge beyond the planet's orbital radius. 
Other mechanisms such as outward migration \citep{Masset2003} could contribute to a gap opening beyond the planet's orbital radius. 
In this paper, and for simplicity, we assume that a planet of a few M$_\mathrm {Jup}$ could exist within the whole extension of the hole/gap region of the five transitional disks in our sample.

\subsection*{\bf Stellar companions}

Binary systems with separations smaller than the outer disk, on the other hand,  are expected to open inner holes that are signi\-ficantly larger than the semi-major axis of the orbit of the binary, while keeping circumprimary disks that dissipate quickly on viscous timescales. In this case, the inner hole or gap is crea\-ted by tidal truncation of the disk which, initially interpreted as transitional, is in fact a circumbinary disk. \cite{Artymowicz1994} investigated the gravi\-tational interaction of an eccentric binary with the circumbinary and circumstellar gaseous disks and analytically computed the approxi\-mate size of the disk gaps as a function of the binary mass ratio and eccentricity. For typical viscous disk parameters and $a$ denoting the semi-major axis of the binary, the inner edge location of the circumbinary disk relative to the center of mass of the binary, or $R_{hole,out}$, varies from $1.7a$ to $3.3a$ for a binary eccentricity increasing from 0 (circular orbit) to 0.75 and assuming a face-on geo\-metry ($i= 0^{\circ}$, $PA= 0^{\circ}$). Therefore, $R_{hole,out}/3.3 < a < R_{hole,out}/1.7$ and the position of the compani\-on $d_S$ varies from $a(1-e)$ to $a (1+e)$, the periapsis and apoapsis of the secondary relative to the primary.  Observations of binary systems inside disk holes suggest that these numbers are approxi\-mately correct (e.g., \citealt{Guenther2007} for CS\,Cha, \citealt{Ireland2008} for Coku\,Tau/4, \citealt{Beust2006} for GG\,Tau, and more recently, \citealt{Huelamo2011} for T\,Cha). 
The circumbinary disk recedes from the binary with increasing eccentricity. For simplicity, we assume the origin of the coordinate system to be coincident with the primary star. 
This will not affect the results for $d_S$  significantly because the clearing is always larger than $a$ and because of   the many uncertainties involved.

In order to compute the expected positions of stellar companions for the stars in our sample, we considered the example in \cite{Artymowicz1994} of a binary star with viscosity parameter $\alpha_\upsilon \sim 10^{-2}$, Reynolds number $R \sim 10^5$ and binary mass parameter $\mu$ = 0.3 for which $R_{hole,out}=1.7a$ for $e = 0$ (model 1), $R_{hole,out}=2.6a$ for $e = 0.25$ (model 2), and $R_{hole,out}=3.3a$ for $e= 0.75$ (model 3). The results are summarized  in Table~3 and are plotted in Fig.~3 as the pink area in the sensitive curves. This area represents the extension within the disk hole/gap where stellar companions could exist according to these models and for a face-on geometry. The upper limit of this region is given by the circular orbit (model 1, $d_S = R_{hole,out}/1.7$)  and the lower limit by the periapsis of the most eccentric orbit (model 3; $d_S = a(1-e) = R_{hole,out}/3.3(1-0.75)$). The corresponding orbital period $P$ was computed assuming Keplerian rotation\footnote {\scriptsize {$P^2 = \frac{4\pi^2}{G(M_\star +~m)}a^3$, where $G$ is the gravitational constant, $M_\star$ the mass of the primary and $m$ the mass of the secondary.}} and $m = 0.5M_\star$ at the location where companions are expected.

\begin{table*}[!thb]

\caption{Predicted binary signatures from the models of \cite{Artymowicz1994}, projection effects caused by the  inclination of the orbit, and hole/gap range where stellar/substellar companions can be ruled out from the NACO images }
\label{table:3}
\bigskip
\scriptsize
\centering
\begin{tabular}{lcccccccccccc}
\hline\hline
\noalign{\smallskip}
Star & R$_{\rm hole,out}$ & \multicolumn{2}{c}{Model 1${\boldmath^{\mathrm{1}}}$} &  \multicolumn{2}{c}{Model 2${\boldmath^{\mathrm{2}}}$} &  \multicolumn{2}{c}{Model 3${\boldmath^{\mathrm{3}}}$} &  $i$ & \multicolumn{2}{c}{Proj. effects${\boldmath^{\mathrm{4}}}$} & $\Delta \rm R_{stellar}{\boldmath^{\mathrm{5}}}$ & $\Delta \rm R_{substellar}{\boldmath^{\mathrm{6}}}$ \\
\cline{3-4}\cline{5-6}\cline{7-8}\cline{10-11}
\noalign{\smallskip}
 &  & $\rm d_S$ & $\rm P$ &  $\rm d_S$ & $\rm P$ &  $\rm d_S$ & $\rm P$ &  & $\rm d_{proj}{\boldmath^{\mathrm{*}}}$ & $\rm d_{proj}{\boldmath^{\mathrm{**}}}$ &  & \\
 \noalign{\smallskip}
 & (AU) & (AU) & (yr) &  (AU) & (yr) &  (AU) & (yr) & (deg) &  (AU) &(AU) & (AU) &(AU) \\
 \noalign{\smallskip}
\hline
\noalign{\smallskip}
DoAr\,21 & 100 &  59 & 276 & $29-48$ & $95-202$ & $7.6-53$ & $13-235$ & $<$ 45 & $42-59$ & $71-100$ & $12.2-100$ & $31-100$   \\
HD\,135344B & 39 & 23 & 69 & $11-19$ & $23-52$ & $3-21$ & $3-60$ & 21& $ 21-23$ & $36-39$ & $14-39$ & $37-39$  \\
HR\,4796A & 69 & 41 & 136 & $20-33$ &  $46-98$ & $5-37$ & $6-116$ &  76 & $10-41$ & $17-69$ & $7.3-69$ & $13-69$ \\ 
T\,Cha & 15 & 8.8 & 17.4 & $4.3-7.2$ &  $6-13$ & $1.1-7.9$ & $0.8-15$ & 60 or 75  & $2.3-8.8$ & $4-15$ & $10.8-15$ & --\\
TW\,Hya &  4.3 & 2.53 & 4.2 & $1.25-2.1$ &  $1.5-3.2$ & $0.33-2.3$ & $0.2-3.7$ & 4.3 & $2.52-2.53$ & $4.29-4.3$ & -- & --  \\
\noalign{\smallskip}
\hline
\end{tabular}
\begin{list} {}{}
\item[${\boldmath^{\mathrm{1}}}$] {\scriptsize Model 1 (circular orbit): $e = 0$, $R_{hole,out}=1.7a$ and the distance of the secondary relative to the primary is $d_S = a =  R_{hole,out}/1.7$ }
\item[${\boldmath^{\mathrm{2}}}$] {\scriptsize Model 2: $e = 0.25$, $R_{hole,out} = 2.6a$ and $a(1-e) < d_S < a(1+e)$ or  $0.29 R_{hole,out} < d_S < 0.48 R_{hole,out}$} 
\item[${\boldmath^{\mathrm{3}}}$]{\scriptsize Model 3: $e = 0.75$, $R_{hole,out} = 3.3a$ and $a(1-e) < d_S < a(1+e)$ or  $0.076 R_{hole,out} < d_S < 0.53 R_{hole,out}$}
\item[${\boldmath^{\mathrm{4}}}$]{\scriptsize Binary coplanar with the disk ($i_S = i_{disk}$), circular orbit ($e = 0$), $x_{proj} = a\,cos\,\theta$, $y_{proj} = a\,sin\,\theta\,cos\,i$ and $d_{proj} = (x_{proj}^2 + y_{proj}^2)^{1/2}$; $d_{proj}(max) =  a$ for $\theta = 0^\circ$ and  $d_{proj}(min) =  a\,cos\,i$ for $\theta = 90^\circ$. Computed for ${\boldmath^{\mathrm{*}}}a = R_{hole,out}/1.7 $ and ${\boldmath^{\mathrm{**}}}a = R_{hole,out}$ }
 \item[${\boldmath^{\mathrm{5}}}$] {\scriptsize Range of disk hole radii, starting at 0\farcs1 from the primary, in which \emph{stellar} binary companions can be ruled out on basis of the VLT/NACO images and using the BT-DUSTY evolutionary models} 
\item[${\boldmath^{\mathrm{6}}}$] {\scriptsize Range of disk hole radii in which \emph{substellar} binary companions can be ruled out from this work} 
\end{list}
\end{table*}

\subsection*{\bf Projection effects}

In this section we evaluate the orbit projection effects on our results on the basis of the disk inclinations $i$ reported in the lite\-ra\-ture for the five sources, assuming the binary is coplanar with the disk. Companions in nearly edge-on systems are likely to be occulted by the primary or by the disk itself. The gap properties also remain unchanged until the disk is nearly edge-on and starts to obscure the central star \citep{Brown2007}.
Assuming a circular orbit for simplicity (we could not access the eccentricity of a detected companion with just one epoch images, anyway) and polar coordinates with the origin of the referential at the primary, the projected orbit can be computed with $x_{proj} = a\,cos\,\theta$ and $y_{proj} = a\,sin\,\theta\,cos\,i$, where $a$ is the semi-major axis of the real orbit, $i$ the inclination and $\theta$ the angle between the position vector and the x-axis. The projected distance between the primary and secondary is determined by $d_{proj} = (x_{proj}^2 + y_{proj}^2)^{1/2}$ at any point in the orbit. The results are listed in Table~3 and, as expected, the projection effects are more significant for the more inclined disks HR\,4796A and T\,Cha. 

In our case, a non-detection of the companion would occur for projected distances shorter than 0\farcs1, the detection limit in radii of the NACO images. Confining the analysis to the first quadrant ($0^{\circ}<\theta<90^{\circ}$), we can determine the angle $\theta$ for which $d_{proj}<0\farcs1$ and predict the percentage of the orbit we could effectively detect with our NACO observations. We performed this exercise for two values of the semi-major axis, $a = R_{hole,out}/1.7$, and $a = R_{hole, out}$. 
The results show that a non-detection of the companion owing to projected distances shorter than 0\farcs1 could \mbox{exist} only for T\,Cha for $a = R_{hole, out} = 15$~AU. At a distance of 108~pc (0\farcs1 $\sim$ 10.8 AU) and for disk inclinations  $i$ = 75$^\circ$ or 60$^\circ$ reported in the literature, a non-detection  could occur if $\theta$ = 45$^\circ$ or 50$^\circ$ (1st~quadrant) or when $ 45-50^\circ < \theta < 225-230^\circ$ and $225-230^\circ < \theta < 315-320^\circ$, for the entire orbit. Therefore, we would access 50 to 55\% or 19.5 to 21.5 yr of a total orbit of 39~yr for a distance of 15~AU from the primary and  $m = 0.5M_\star$.  If $a = R_{hole}/1.7 = 8.8$~AU, the NACO images are no longer sensitive to the distances involved. This is also the case for TW\,Hya, for which 0\farcs1 $\sim$ 5.5~AU and $R_{hole,out} = 4.3$~AU. 

\section{Discussion: hole/gap formation mechanisms}
\label{discussion}

The origin of transitional disks is still under debate with
se\-ve\-ral theories developed to explain inner opacity holes and gaps in
protoplane\-tary disks, including  i) magneto-rotational instabili\-ty \citep[MRI;][]{Chiang2007}, 
ii) photoevaporation by the central star \citep{Clarke2001,Alexander2006,Gorti2009}, iii) dust removal by coagu\-lation \citep{Tanaka2005, Dullemond2005} and iv) dynamical clearing by unresolved companions, both stars \citep{Artymowicz1994}, and giant planets \citep{Quillen2004,Varniere2006b}. 
 
The MRI mechanism operates in ionized gas and hence, at all disk surface layers ionized by direct X-rays from the central star, including the inner disk wall. Although this still allows accretion onto the star, MRI cannot explain the remnant optically thick disk observed in ``pre-transitional" disks. 
Current photoevaporation models require a maximum total disk mass of $\sim$~5~M$_{\rm Jup}$ and negligible accretion rate, unless the disk is un\-usually inviscid or large ($>$ 100~AU). Extreme-UV (EUV), far-UV (FUV) photons and X-rays from the central star heat and ionize the circumstellar hydrogen, generating a disk evaporative wind beyond some critical radius. At disk early-evolutionary stages, the accretion rate dominates over the evaporation rate. For photoevaporation to hold, the accretion rate has to fall below the wind rate and, once this happens, the photoevaporative wind can open a gap in the gas disk. The dust in the inner disk is then rapidly removed (on a viscous timescale of typically less than $10^5$~yr), for\-ming the inner hole, and the entire disk dissipates very quickly through photoevapo\-ration. Computed disk lifetimes are $\sim$ 4 Myr for M$_\star \leq 3$ M$_\odot$ and less than 1 Myr for more massive stars \citep{Gorti2009}. The inner disk drains away entirely before the gap can grow beyond 1-10~AU, and therefore  photoevaporation cannot explain  the inner dust disks and large inner holes observed in many transitional objects.
Grain growth can also produce opa\-city holes but is a strong function of the radius. It is more efficient at disk small radii where the surface density is higher and the dynamical timescales are shorter and do not affect accretion (of the gas). Unfortunately, is not possible to detect grains larger than a few millimeter and, therefore, this mechanism was not observationally confirmed in the last years. And finally, as discussed in section \ref{sec:orbit}, stellar/substellar companions or giant planets of 1 to a few M$_{\rm Jup}$ can open gaps and inner holes through tidal interaction with the disk. 
While a stellar companion is expected to almost suppress  accretion, an embedded planet of $\sim$ 1 M$_{\rm Jup}$ has only a modest effect on the disk accretion rate, reducing it by a factor of $\leq$ 10 inside the planet's orbit compared to that from outside, and has little effect on the total disk mass \citep{Lubow2006}. And, for a given disk mass,  the more massive the planet and the less viscous the disk, the lower the accretion rate \citep{Alexander2007}. \cite{Lubow1999} proposed that a planet of 10 M$_{\rm Jup}$ or more could halt stellar accretion. Nevertheless, a few observed close stellar/substellar binaries have been reported to be accreting: DQ\,Tau \citep{Carr2001}, CS\,Cha \citep{Espaillat2007} and, more recently, T\,Cha \citep{Huelamo2011, Olofsson2011}. Factors other than the presence and mass of a close companion must determine accretion onto the star and hence, the accretion rate alone cannot be used to infer or discard the presence of potential com\-pa\-nions. The viscosity and scale height of the disk and eccentricity of the orbit may contribute to maintain the accretion across the disk gap/hole even for stellar companions \citep{Artymowicz1996}.   

\cite{Alexander2007} proposed a simple observational diagnostic to distinguish between photoevaporation and giant planet-induced inner holes that relies on the relative va\-lues of the the total disk mass (M$_{\mathrm {disk}}$) and stellar accretion rate ($\dot M_{\rm acc}$). More recently, \cite{Cieza2010} used optical-to-millimeter observations of a sample of 26 transitional disks in the Ophiuchus  molecular cloud (d $\sim$ 125 pc) to derive the disk parameters and investigate the mechanisms that potentially creat\-e their inner holes. In addition to the disk mass and accretion rate, the authors also  consider the fractional disk lumino\-si\-ty ($L_{\mathrm {disk}}/L_\star$) and parameters from the SED. According to their criteria, grain-growth-dominated disks have high-accretion rates ($\dot M_{\rm acc} \geq10^{-8}$~M$_\odot \mathrm {yr}^{-1}$) and large disk masses because  this mecha\-nism mostly operates in the disk inner region, whereas most of the mass resides at larger radii. Photoevaporating or photoevaporated disks should have a small disk mass (M$_{\mathrm {disk}} <  2.5~M_{\mathrm {Jup}}$) and negligible accretion ($ < 10^{-11}$ M$_\odot \mathrm {yr}^{-1}$), and if in addition  the fractional disk luminosity $L_{\mathrm {disk}}/L_\star < 10^{-3}$, the disk could be already in the debris evolutionary stage regardless the origin of the inner hole.  Good candidates for ongoing giant planet formation are disks with large mass (1.5 - 11 $M_{\mathrm {Jup}}$) and low accretion rate ($\sim 10^{-10} - 10^{-9}$ M$_\odot \mathrm {yr}^{-1}$), while disks with small mass ($< 1.5 M_{\mathrm {Jup}}$) and higher accretion ($10^{-9.3} - 10^{-7.3}$~M$_\odot \mathrm {yr}^{-1}$) could have formed a giant planet in a recent past because most of the disk mass has already been depleted. 
Close binaries are ruled out on the basis of the \cite{Artymowicz1994} mo\-dels ($R_{hole,out} \sim 2a$). 

Following these criteria and using the disk parameters in Table~2, we tentatively classify the five transitional disks in our sample on the basis of the mechanisms that potentially create their inner opacity hole/gap. DoAr\,21 was classified as a possible circumbinary or debris disk by \cite{Cieza2010} (source 12 in their catalog). HD\,135344B, T\,Cha, and TW\,Hya are good candidates for {\it ongoing} giant planet-formation because of their large disk masses and modest accretion rates. HR\,4796A could be a photoevaporated, debris, or a planetary disk owing to its very low mass, dust ring structure and no reported accretion rate. Nevertheless, this classification is just a {\it best guess} given the available data and the uncertainty associated to all parameters, both observational or derived from the models. Additionally, the hole/gap clearing mechanisms are not all mutually exclusive. Two or more mechanisms can operate at the same time (for example, grain growth and MRI, or planet formation and photoevaporation) or sequentially (once the gap has formed by one mechanism, others can rapidly remove the inner disk ), contributing to the present observed transitional evolutionary state of the disk. Transitional disks seem to represent a highly hete\-rogeneous group of objects exhibiting a wide range of masses, accretion rates, and SED morphologies although, overall, they tend to have smaller masses and lower accretion rates than disks without holes \citep{Cieza2010}. To conclude, disk dissipation does not seem to follow a standard evolutionary path, but instead many different paths are possible to which different mechanisms contribute. 

The VLT/NACO images presented here provide an observational input to the last proposed mechanism by ruling out the presence of stellar companions within part  (beyond 0\farcs1) of the inner hole/gap region of four disks in our sample and down to the planetary mass regime for DoAr\,21 and to the substellar mass regime for HD\,135344B and HR\,4796A. The exception is TW\,Hya, for which the NACO images cannot resolve the inner hole of just 4.3~AU, or 78~mas for a distance of 55~pc. The results in Fig.~3 (pink bars in the plots) and Table~3 show that our observations are sensitive to part of the region where binaries are expected to exist from the models of \cite{Artymowicz1994} for HD\,135344B and almost this entire region in the case of HR\,4796A and DoAr\,21. With the proviso of possible occultations, we do not report any detection of companions beyond 0\farcs1 from the primary for the five transitional disks.
Therefore, the results in this paper further cons\-train the suite of potential explanations for their origin and toward the exciting possibili\-ty that giant planets are currently forming inside their inner holes/gaps.
Nevertheless, the fraction of transitional disks that are in fact tight binaries still remains to be established. Common adaptive optics observations (on 8 to 10m class telescopes)  cannot cons\-train the inner 0\farcs1 region from the primary, or radii of less than 5 to 15~AU for most transitional disks. Additional surveys \mbox{using} different techniques such as sparse aperture masking (SAM) with adaptive optics instruments (the one leading to the detection of the potential substellar companion of T\,Cha  reported in \citealt{Huelamo2011}) and radial velocity studies, should be performed to probe the inner most regions and firmly establish the fraction of transitional disks that are in fact close binaries. Finally, the Atacama Large Millimeter Array (ALMA) with its 1-3~AU spatial resolution and high sensitivity will be capable of resolving young planetary systems, detecting and determi\-ning the properties of close companions (both planets and stars) and of the dust and gas through out the disk. Planet and inner hole/gap formation scenarios will be confirmed or ruled out.

The following sections summarize the results regarding the presence of potential companions for each individual source from previous studies and this work. 

\subsection*{DoAr 21}

\cite{Jensen2009} summarizes the properties of this object. Gemini sub-arcsecond resolution 9-18 $\mu$m images show little or no excess mid-IR emission within 100~AU from the star. But beyond that the emission extends over several arcsec (or hundreds of AU) and is quite asymmetric, showing a half-ring structure of higher emission at 1\farcs1 from the star to the NW. This feature is bright in PAHs emission at 8.6 and 11.3 $\mu$m, H$_2$  emission at 2.12~$\mu$m (the 1$-$0 S(1) ro-vibrational line) and conti\-nuum emission at  18 $\mu$m, and very different from what is expected from a disk.  DoAr\,21 is a particular object different from other pre-main sequence stars (PMS) because it is very young,  a very strong emitter of (hard) X-rays and shows no accretion. Based on the estimated   FUV and X-ray flux at the position of the ring ($\sim$130~AU) and the proper motion of DoAr\,21 ($9.5 \pm 1.3$~kms$^{-1}$ westward),  \cite{Jensen2009}  proposed this emission to be associated with a small photodissociation region (PDR) resul\-ting from the interaction of the radiation field with denser material from the Ophiuchus cloud in the vicinities of the star and not with a circumstellar (or circumbinary) disk. The half-arc shape would form because the material is swept up by the stellar motion. The H$_2$ emission was spatially resolved with the AO-assisted IFU instrument SINFONI at the VLT as an arc of material circling the star at radii between 0\farcs6 (73~AU) and 1\farcs8 (220~AU) and position angles of $ -40^\circ$ to $170^\circ$ by M.~Hogerheijde et al. (to be submitted), who proposed several explanations for the asymmetric emission. The derived H$_2$ line ratios were consistent with thermally excited gas by X-rays (T = 1000-2000~K) and not with FUV fluorescence. The most likely scenarios include illumination of unrelated cloud material or material resultant from the collision between two (proto)planets and the migration of a~substellar/planetary companion to the ring region (73-220~AU) that could disrupt a previously stable disk. The last scenario was ruled out by their SINFONI images down to a mass limit of 5~M$_{\mathrm {Jup}}$, by assuming an age of 0.3 Myr for DoAr\,21. However, an unseen planet of a few M$_{\mathrm {Jup}}$ could still explain the observations. Future epoch observations  are necessary to determine if this material is orbiting the star (in a disk) or if is tra\-ve\-ling through the cloud together with DoAr\,21, confirming the PDR scenario.

VLBA observations by \cite{Loinard2008} show  DoAr\,21 as a binary with 5 mas (0.6~AU) projected separation  and estimated semi-major axis of 1-2 ~AU, but no information is given on the mass or optical luminosity ratio. Assuming it to be an equal-mass binary, evolutionary  tracks predict for each star a mass of $\sim$ 1.8 M$_\odot$ ($\sim$ 3.6 M$_\odot$ for the binary) and an age of $\sim 8 \times 10^5$~yr. Nevertheless, such a close companion is beyond the resolution of NACO. 

As expected, the H$_2$ emission was not reco\-ve\-red in our VLT/NACO 2.12~$\mu$m  images (the width of the H$_2$ line in this source is of 2$\times$10$^{-4}$ $\mu$m and was detected with an integration time of 7 seconds and a S/N of $\sim$10). Taking into account the width of 0.022 $\mu$m of the NB\_2.12 NACO filter, one would have needed integration times 220 times longer  (i.e., 25 min) to achieve a similar S/N with our observations. Despite failing to detect  off-source emission at 2.12 $\mu$m,  the NACO images confirm the absence of substellar/planetary companions in the arc region down to a mass of 15 M$_{\mathrm {Jup}}$ at 73~AU and 6~M$_{\mathrm {Jup}}$ at 192~AU. Our NACO images conservatively exclude substellar objects (m~$< 75$ M$_{\mathrm {Jup}}$) at radii larger than $\sim$ 31~AU (0\farcs25)  and of stellar companions within the extension of the circumstellar environment probed with these observations  (r~$\gtrsim$ 0\farcs1 or 12~AU). The 5$\sigma$ sensitivity at 0\farcs1 is 3.1 mag or 0.19 M$_\odot$ for an assumed age of 1 Myr.  

 \cite{Cieza2010}  collected nearly diffraction VLT/NACO/S13 images in the $J$- and $K_s$-bands (FWHM of 60--70 mas) from April to September 2009 of several transitional disks, including DoAr\,21.  A visual inspection of those images revealed no stellar companions beyond $\sim$~8~AU from the primary, which agrees with our results. Their detection limits at 0\farcs1 separation, estimated from the 5$\sigma$ noise in the PSF subtracted images, have a median value of 3.1 mag corresponding to a flux ratio of 17. These authors classified DoAr\,21 as a possible circumbinary or debris disk based on the disk parameters and no observed accretion (see discussion above). As proposed also by \cite{Jensen2009}, the DoAr\,21 disk must have almost entirely dissipated either through binary formation or photoevaporation. This scenario is indeed possible. An equal-mass eccentric binary with semi-major axis of 1-2~AU would tidally clear out an inner hole of 3 to 6~AU. Once the gap was opened, the high FUV and X-ray luminosity of the binary with mass of $\sim$ 3.6~M$_\odot$  would photoevaporate the disk on a timescale of $\sim 10^5-10^6$ yr \citep{Gorti2009}. 
 
The NACO images of DoAr\,21 are the only ones from the sample of five disks that are sensitive enough for a detection of objects less massive than $\sim$~13~M$_{\rm Jup}$ that is, potential planets or low-mass brown dwarfs within the inner hole region of the disk and at radii larger than $\sim$~76~AU (0\farcs63).

\subsection*{HD\,135344B or SAO~206462}

Direct imaging at 880 $\mu$m with the Submillimeter Array (SMA) by \cite{Brown2009} spatially resolved the disk and an inner hole extending to a radius of 39~AU, for a distance of 140~pc to HD\,135344B, in agreement with the value of 45~AU derived from the SED best-fit model \citep{Brown2007}. 
This star has a wide companion at 20\farcs4 \citep{Coulson1995}. HD\,135344B has been observed by the HST in coronographic mode with NICMOS (1.1 and 1.6 $\mu$m) and STIS and no companions were found \citep{Grady2005,Grady2009}. The 5$\sigma$ detection sen\-si\-ti\-vity limits show the non-existence of  stellar compa\-ni\-ons within $\sim$~0.1 to 1~arcsec ($\sim$ 14 to 140 AU) from the star.  \cite{Pontoppidan2008} detected CO emission  at 4.7~$\mu$m extending from 0.3 to 15~AU from the primary with CRIRES at the VLT, arguing against stellar clearing, but in favor of  a pla\-ne\-tary mass companion orbiting at 10-20~AU. Spectroastrometry by \cite{Baines2006} also excludes stellar companions farther out than 0.42 AU. All these studies cannot exclude substellar objects within the disk (160~AU). The current NACO observations extend the previous results by excluding brown-dwarf-mass companions at orbits larger than $\sim$~37~AU (0\farcs27) and down to a mass limit of $\sim$ 19~M$_{\rm Jup}$ (8~mag in contrast) at 221~AU. At 0\farcs1 they are sensitive to stars more massive than 0.22 M$_\odot$. 

\subsection*{HR\,4796A}

 \cite{Schneider2009} obtained high spatial resolution ($\sim$~70~mas) optical coronographic images of the HR\,4796A debris dust ring with the Space Telescope Imaging Spectrograph (STIS) resolving emission from $\sim$ $69-87$~AU ($0\farcs95-1\farcs2$) in radii (see Fig.7 in their paper) and a displacement of 1.4~AU between the disk center and the star, measured along the ring major axis. These values are in accordance with previous observations at 1.1 and 1.6 $\mu$m  with the HST/NICMOS instrument \citep{Schneider1999}. The hole outer radius was first inferred from the SED by \cite{Jura1995} to be from 40 to 200~AU. HR\,4796A is a visual binary having a M2.5-dwarf companion at  7\farcs7 or 560~AU \citep{Jura1993}. More recently, a tertiary wide companion ($\sim$ 3$\arcmin$separation) was discovered from its X-ray emission and is a M4.5 spectral-type star \citep{Kastner2008}. 
No other companions have been detected in $H$- and $K_s$-band coronographic NACO imaging at projected separations in the range 0\farcs1 to 10$''$ \citep{Chauvin2010} and there are no radial velocity measurements for this source. \cite{Wyatt1999} concluded that a planet more massive than 10 M$_{\rm Earth}$ close to the inner edge of the ring  would be required to explain the asymmetry seen in the dust emission. Such a low-mass object could not be detected with the observations presented here. Nevertheless, the NACO images discard the existence of stellar compa\-nions within the disk hole at radii larger than $\sim$ 7~AU (0\farcs1) and more massive than 0.24 M$_\odot$, and of substellar objects outwards of $\sim$~13~AU (0\farcs18). The mass detection limit at the hole outer radius (69~AU) and inside the dust ring (up to 87~AU) is 17~M$_{\rm Jup}$. 

\subsection*{T\,Cha}
\label{sec: tcha}

Radial velocity (RV) observations were performed for this source and the results are presented in \cite{Schisano2009}. The authors report a radial velocity variation of 10 kms$^{-1}$, albeit non-periodic. No companions for T\,Cha were detected at projected separations of 0\farcs1 to 10$''$ in diffraction-limited $H$- and $K_s$-band  NACO images with Lyot-coronography  \citep{Chauvin2010}.  Recently, \cite{Huelamo2011} reported the presence of a potential substellar companion at a separation of $62 \pm 7$~mas ($6.7 \pm 1$~AU) from T\,Cha detected in the $L'$-band (contrast $\Delta L' = 5.1 \pm 0.2$~mag) with NACO and using sparse aperture masking (SAM). The object was not detected in the authors'  $K_s$-band data up to a 3$\sigma$ contrast limit of 5.2~mag. The corresponding 5$\sigma$ contrast limit is $\sim$  4.65~mag or 0.12~M$_\odot$ for the BT-DUSTY model adopted here and is indicated in Fig.~3 as the blue vertical bar in the T\,Cha sensitivity plot. A lower limit of~50~M$_{\rm Jup}$ was chosen following Fig.~4 in \cite{Huelamo2011} paper. 
This detection is well within the T\,Cha disk hole/gap and inside the binary  range predicted by the models of \cite{Artymowicz1994}\footnote{\scriptsize Assuming the companion is coplanar with the disk, which has an inclination of 60$^\circ$ to 75$^\circ$ reported in the literature, and a circular orbit, we can put some constraints on the orbit of the potential companion by considering the two extreme cases: 1) $\theta$ = 0$^\circ$ and the distance primary-secondary  equals the binary semi-major axis, $d_s = a = 6.7$~AU, and 2) $\theta$ = 90$^\circ$ and $d_s = acos\,i = 6.7$~AU. In 1) the potential companion is at its largest projected separation at the moment and will reach the shortest separation of 1.7 to 3.3~AU when $\theta$ = 90$^\circ$ for an inclination varying between 60$^\circ$ and 75$^\circ$. The orbital period varies between 13.6~yr (for $m = 0.12$~M$_\odot$) and 14~yr (for $m = 50$~M$_{\rm Jup}$). In 2) the companion is at its shortest projected separation in the present and the semi-major axis of the orbit $a$ can be derived assuming an inclination of the orbit $i$: for $i = 60^\circ$, $a = 13.4$~AU and for $i = 75^\circ$, $a = 26$~AU. The orbital period varies between $\sim$ 39~yr ($i = 60^\circ$) and $\sim$ 105~yr ($i = 75^\circ$). Future epoch observations will better constrain the orbit of the potential companion by providing the inclination and the eccentricity. }.
 
Although we could not have detected such an inner low-mass object, these observations support ours and the \cite{Chauvin2010}  results by ruling out stellar companions outward of 0\farcs1 from the primary. 
The NACO images of T\,Cha exclude stellar companions brighter than 2.2~mag in contrast (0.29~M$_\odot$) at a projected separation of 0\farcs1 to 2.9~mag (0.25~M$_\odot$) at 15~AU, or within the disk inner hole. They are not sensitive enough for the detection of substellar binaries within the whole extension of the disk.  The maximum attai\-nable contrast is 4.7 mag, correspon\-ding to a 5$\sigma$ sensitivity of 0.12~M$_\odot$. 

\subsection*{TW\,Hya}

 Detailed modeling of the SED of the star predicted a disk hole/gap outer radius of $\sim$ 4~AU \citep{Calvet2002}  later on confirmed with Very Large Array (VLA) observations at 7 mm by \cite{Hughes2007}. The data resolved the hole in the dust disk and revealed a bright inner rim associated with the disk edge being directly illuminated by the star.
\cite{Setiawan2008} reported a 10  M$_{\rm Jup}$ companion at 0.04~AU from the star through RV observations in the optical regime. \cite{Huelamo2008} carried out near-IR high-resolution observations of the source with the CRyogenic high-resolution InfraRed Echelle Spectrograph (CRIRES) mounted on the VLT and found no trace of this companion. Spectro-astrometric imaging of the 4.7 $\mu$m rovibrational lines of CO by \cite{Pontoppidan2008} detected gas in the TW\,Hya disk hole between 0.1 and 1.5~AU.  Based on this result, the authors suggest that the presence of a $\sim$16 M$_{\rm Jup}$ body could be clearing the dust  and rule out  photoevaporation as the hole formation mechanism. They further explain that if the 4~AU hole is created by a planet, it should be placed in an orbit of $1- 2$~AU. The present NACO observations cannot resolve these inner disk regions. 

Direct coronographic  imaging of the disk in scattered light at 1.1 and 1.6~$\mu$m with the HST/NICMOS instrument  was collected by \cite{Weinberger2002}. The scattering profile indicates a flared disk extending from $\sim$20 to 230~AU. No point sources were detected with a 3$\sigma$ significance from $\sim$ 0\farcs54 to 4$''$ (or 30 to 220 AU) down to a mass of $\sim$ 10 M$_{\rm Jup}$ at $\sim$ 0\farcs63 (35 AU) and 3~M$_{\rm Jup}$ at 1\farcs52 (85 AU). Additionally, TW\,Hya was included in an adaptive optics imaging survey with the Keck~II 10m teles\-cope by \cite{Brandeker2003}.  The star was observed in the $H$-band and  reached a 5$\sigma$ contrast sensitivity of $\sim$ 2 mag at a separation of 0\farcs1. No companions were detected between 0\farcs042 (the diffraction limit in the $H$-band) and 1\farcs6,  or from 2.3 to 88~AU.  

The results derived  here from the NACO imaging further complement and support the previous studies by excluding stellar binaries more massive than 0.11~M$_\odot$ (3~mag in contrast) outward of $\sim$~5.5~AU from the primary (the inner detection limit of the observations or 0\farcs1), brown-dwarf-mass companions outward of  7~AU (0\farcs13) and planetary mass companions outward of $\sim$~13~AU (0\farcs24). We reach the maximum sensitivity of 7~M$_{\rm Jup}$, or 8~mag in contrast from 87~AU onward.  

A recent paper by \cite{Vacca2011} based on near-infrared spectroscopic data revises the spectral type and age of  TW\,Hya to M2.5 (instead of K7) and $\sim$ 3~Myr (instead of $8-10$~Myr). Under the assumption of a younger age, the NACO images are now sensitive to substellar mass companions within the whole disk region probed (0\farcs1 to 7\farcs0) and to planetary-mass objects beyond $\sim$ 10~AU (0\farcs18).


\section{Conclusions}
\label{conclusions}

This paper addresses one possible origin for inner holes or  cleared gaps in protoplanetary disks. These holes/gaps  are seen as potential signposts of planet formation but other explanations are possible. Here we investigate the mechanism of tidal truncation by close binaries as the possible or alternate explanation for the origin of the inner holes/gaps observed in five transitional disks.
The conclusions are that: 

 \begin{enumerate}
      \item With the proviso of possible unfavorable projections, the VLT/NACO images rule out the presence of binary compa\-nions from 0\farcs1 to $\sim$~7\arcsec~in all five transitional disks and down to a 5$\sigma$ detection limit 
      of  2.2--4.9 mag in contrast for the inner region (at 0\farcs1) and 7.7--9.3~mag (4.7 mag in the case of T\,Cha) for the outer region (beyond 1\farcs0).
 
      \item  For the transitional disks DoAr\,21, HD\,135344B, HR\,4796A, and T\,Cha, the 
      VLT/NACO observations resolve part of the inner hole/gap region of the disk and can potentially 
      discard binaries down to a certain mass limit as the main explanation for the inner opacity cavities. For these objects, our results reduce the suite of potential mechanisms that create the holes to closer or lower-mass companions, giant planet formation, efficient grain growth or photoevaporation (for DoAr\,21 and HR\,4796A).
      
 \end{enumerate}

\begin{acknowledgements}
The authors thank the anonymous referee for important comments, which improved this paper, and to C\'atia Cardoso for fruitful discussion on binary dynamics.
Support in the preparation of this publication was provided by the ESA Science Faculty.
 This publication makes use of data
products from the Two Micron All Sky Survey, which is a joint project
of the University of Massachusetts and the Infrared Processing and
Analysis Center/California Institute of Technology, funded by the
National Aeronautics and Space Administration and the National Science
Foundation. This work has made use of the Vizier Service provided by
the Centre de Donnes Astronomiques de Strasbourg, France
(Ochsenbein et al. 2000).  \emph{Image Reduction and Analysis
  Facility} (IRAF) is distributed by the National Optical Astronomy
Observatory (NOAO), which is operated by the Association of
Universities for Research in Astronomy Inc. (AURA), under cooperative
agreement with the National Science Foundation (NSF).
\end{acknowledgements}

\bibliographystyle{aa}
\bibliography{biblio}

\end{document}